\documentclass[twocolumn, switch]{article} % Method A for two-column formatting

\usepackage{preprint}

\usepackage{amsmath, amsthm, amssymb, amsfonts}

\usepackage[numbers,square]{natbib}
\usepackage[utf8]{inputenc}	% allow utf-8 input
\usepackage[T1]{fontenc}	% use 8-bit T1 fonts
\usepackage{xcolor}		% colors for hyperlinks
\usepackage[colorlinks = true,
            linkcolor = purple,
            urlcolor  = blue,
            citecolor = cyan,
            anchorcolor = black]{hyperref}	% Color links to references, figures, etc.
\usepackage{booktabs} 		% professional-quality tables
\usepackage{nicefrac}		% compact symbols for 1/2, etc.
\usepackage{microtype}		% microtypography
\usepackage{lineno}		% Line numbers
\usepackage{float}			% Allows for figures within multicol

\usepackage{lipsum}		%  Filler text

\usepackage{newfloat}
\DeclareFloatingEnvironment[name={Supplementary Figure}]{suppfigure}
\usepackage{sidecap}
\sidecaptionvpos{figure}{c}

\usepackage{titlesec}
\titlespacing\section{0pt}{12pt plus 3pt minus 3pt}{1pt plus 1pt minus 1pt}
\titlespacing\subsection{0pt}{10pt plus 3pt minus 3pt}{1pt plus 1pt minus 1pt}
\titlespacing\subsubsection{0pt}{8pt plus 3pt minus 3pt}{1pt plus 1pt minus 1pt}

\usepackage{tikz,xcolor,hyperref}

\definecolor{lime}{HTML}{A6CE39}
\DeclareRobustCommand{\orcidicon}{
	\begin{tikzpicture}
	\draw[lime, fill=lime] (0,0) 
	circle [radius=0.16] 
	node[white] {{\fontfamily{qag}\selectfont \tiny ID}};
	\draw[white, fill=white] (-0.0625,0.095) 
	circle [radius=0.007];
	\end{tikzpicture}
	\hspace{-2mm}
}
\foreach \x in {A, ..., Z}{\expandafter\xdef\csname orcid\x\endcsname{\noexpand\href{https://orcid.org/\csname orcidauthor\x\endcsname}
			{\noexpand\orcidicon}}
}
\title{The Flux Operator}

\usepackage{authblk}

\author[1\thanks{\tt{sochat1@llnl.gov}}]{Vanessa Sochat\orcidA}
\author[2]{Aldo Culquicondor}
\author[2]{Antonio Ojea}
\author[1]{Daniel Milroy\orcidB{}}
\affil[1]{Lawrence Livermore National Laboratory}
\affil[1]{Google}

\begin{document}

\twocolumn[ % Method A for two-column formatting
  \begin{@twocolumnfalse} % Method A for two-column formatting
  
\maketitle

% \section{Abstract}
\begin{abstract}

Converged computing brings together the best of both worlds for high performance computing (HPC) and cloud-native communities. In fact, the economic impact of cloud-computing, and need for portability, flexibility, and manageability make it not important, but inevitable. Navigating this uncharted territory requires not just innovation in the technology space, but also effort toward collaboration and sharing of ideas. With these goals in mind, this work first tackles the central component of running batch workflows, whether in cloud or HPC:  the workload manager. For cloud, Kubernetes has become the de facto tool for this kind of batch orchestration. For HPC, the next-generation HPC workload manager Flux Framework is analogous -- combining fully hierarchical resource management and graph-based scheduling to support intelligent scheduling and job management. Convergence of these managers would mean the implementation of Flux inside of Kubernetes, allowing for hierarchical resource management and scheduling that scales impressively without burdening the Kubernetes scheduler itself. This paper introduces the Flux Operator -- an on-demand HPC workload manager that is easily deployed in Kubernetes. The work here highlights design decisions, mapping of components between environments, experimental features, and shares the results of experiments that compare performance with an equivalent operator in the space, the MPI Operator. Finally, discussion closes with a review of challenges remaining, and hopes for the future for improved technological innovation and collaboration.

\end{abstract}

%\keywords{First keyword \and Second keyword \and More} % (optional)
\vspace{0.35cm}

  \end{@twocolumnfalse} % Method A for two-column formatting
] % Method A for two-column formatting

%\begin{multicols}{2} % Method B for two-column formatting (doesn't play well with line numbers), comment out if using method A

% High level TODOs from v
% 1. Be consistent in use of "scheduler" vs "workload manager"
% 2. I asked Evan in slack. Do the clouds publish how many unique users use their services.
% 3. We need to be sure that we are consistently targeting an audience, both for the readership and "who cares about this worki in the paper" arguably there are "the developer user and also people who are creating large, interdisciplinary workflows that recognize cloud is important" (both) in there.
% 4. Dan bad words to search for: "obvious" "easy" "layman" "we" "our"

%%%%%%%%%%%%%%%  Main text   %%%%%%%%%%%%%%%
% \linenumbers

% \section{Introduction}
\section{Introduction}
\label{sec:introduction}

Portability, manageability, and modularity of complex, heterogeneous workflows is becoming increasingly important for high performance computing (HPC). In particular, the need for workflows to be extended to cloud environments is a key component of collaboration across an increasingly diverse set of computational resources, and a likely solution for ``green computing" to ensure energy efficiency and optimal usage of shared resources \cite{bharany2022systematic}. Other demands for flexibility of compute include the increasing use of internet of things ``IoT'' remote devices to conduct research \cite{sadeeq2021iot,zhu2023quakeflow}, an explosion of hardware available on cloud platforms \cite{Thompson2021-ww,reuther2022ai}, and the dynamic addition of external resources \cite{jena2022high}. A powerful demonstration of need also comes from a series of events organized by the European Commission \cite{digital-autonomy-computing-continuum} to assemble experts for discussion on innovation in the ``computing continuum," citing a strong need for flexibility  for distributed systems, green and dynamic technologies, and an emphasis on open source software. The discussion continues with workshops \cite{workshops-cloud-computing} emphasizing the importance of shaping Europe's digital future. Given this landscape, any entity involved in the business of scaled computing will fall behind if these technological needs are not prioritized \cite{Thompson2021-ww}.

In cloud computing communities, machine learning workloads are also becoming increasingly important \cite{doi:10.1080/17460441.2021.1932812,george2022end,kreuzberger2023machine,liu2022scanflow}, and the cloud container orchestration technology Kubernetes \cite{noauthor_2023-vz} has become the de facto standard for orchestration of these workflows. As of June of 2023, the Kubernetes project had approximately 74,000 contributors, making it the second largest open source project ever after Linux, and the ``most widely used container orchestration platform in existence'' \cite{noauthor_2023-vz}. Outside of the academic community it is the chosen platform for orchestration, being used at over 70\% of Fortune 500 companies \cite{noauthor_2023-vz}. In recent years \cite{noauthor_undated-qv}, the growing need for supporting batch workflows \cite{kubernetes-evolution-batch-powerhouse-article} has led to the batch working group. This group works on the design and implementation of application programming interfaces (APIs) to enable cloud-native batch workflows and jobs, and provides an interesting transition of Kubernetes from primarily a stateless, service-oriented architecture to one that can support states and a desired completion of work. The first stable release \cite{kubernetes-126-jobs-aldo} of the Job controller marked an unofficial declaration of Kubernetes supporting what, at face value, looked like more traditional workflows from HPC. This development made the idea of deploying one of the world's top supercomputers in Kubernetes an achievable goal \cite{noauthor_2022-te}, and the needs of the cloud computing communities overlapped better with the needs of HPC than ever before. 

In the high performance computing space batch processing has a long history, and consequently the community has deep expertise \cite{history-batch-processing}. The need to embrace traditionally more cloud-like features arguably came down to the demands of the workloads themselves. While a traditional HPC workload might be embarrassingly parallel, meaning running equivalent, scoped tasks across a homogeneous set of resources concurrently, modern workflows include the gamut from simulation to batch processing to artificial intelligence (AI) and services. A standard workflow is no longer a single run that writes to a shared filesystem, but rather an assortment of tasks that vary in their needs for hardware, storage, and running times. Modern workflows are typically provided via directed acyclic graphs (DAGs) that not only indicate direction or order, but also utilization of entirely different architectures, services, and virtualization technologies.  Indeed, the high performance computing community needed the same portability, flexibility, and automation for these workflows afforded by cloud, and to span both applications and services. 

It would be in the best interest of cloud communities to learn from and take on the best technological innovations from HPC, and vice versa. Thus, this landscape with overlapping interests was a spark for collaboration, and the time was right for the convergence of not just these communities, but the technologies themselves. This collaboration, or the convincing of one community to engage with the other, is arguably more challenging than the development work itself. When approaching those on the HPC side, discussions that suggest using cloud very quickly turn to the matter of performance, costs \cite{munhoz2022strategies,munhoz2022hpc} and security \cite{jangjou2022comprehensive}. Approaching the cloud side, there is often lack of understanding for how high performance computing technologies might be useful or needed. In the case of the first, a simple solution that resolves many concerns is the often forgotten reality that cloud setups can be public or private. A technology such as Kubernetes could be deployed on-premises. In the case of the second, arguably the cloud computing community needs convincing that they too can benefit from the adoption of HPC technologies. Increasing performance and efficiency by using techniques from HPC, and providing better transparency of underlying resources by way of low-level performance analysis, would both lower costs and time to completion. A solution that falls in the middle would likely bring together the best of both worlds, but could come with compromises such as paying a performance penalty for flexibility. Ideally, a converged approach would be able to most effectively use hardware and improve performance, with increased flexibility offsetting any potential compromises for said performance.

Anticipating a desired future where cloud and high performance computing communities are collaborating and developing solutions together, the challenges stated above can be inspected to understand what is needed for both sides. First, a solid demonstration is needed that there are benefits for both cloud and HPC to take on attributes of the other side. For HPC, this means more modularity, portability, and automation. For cloud, this means more performant workflows, efficient use of hardware, schedulers, and communication protocols that span networking, applications, and storage.  Secondly, examples of such technologies must be prototyped that can bring together the best of both worlds - the performance of HPC and the flexibility of clouds. This vision, or the space of technologies that exists between cloud and HPC can be described with the term ``converged computing" \cite{milroy2022one,misale2021towards}. In a converged computing landscape of the future not only will technologies from traditionally disparate communities be brought together, but traditionally disparate communities will be united culturally to identify shared goals and foster deeper, more meaningful collaborations.

While many areas of work can be tackled, it was logical to start with a workflow manager analogous to Kubernetes in the high performance computing community, with the common use case of running simulations or machine learning tasks. The Flux Framework, a novel hierarchical framework for resource management and scheduling, provides similar abstractions that parallel those in Kubernetes, including modularity, well-defined developer and user interfaces, and an ability to integrate and manage different types of resources. It stands out from other resource managers because of its ability to manage the exploding complexity of modern workflows described previously. To start with modularity, in the same way that several components are combined to create Kubernetes \cite{kooberdoober-components}, components from Flux Framework \cite{flux-projects} are assembled together to manifest in a workload manager that is referred to simply as ``Flux." This modularity might mean, for example, that a component of Flux could be used in Kubernetes, or vice versa. For developer interfaces, arguably a core ingredient of convergence is having common programming languages or bindings. Kubernetes, as it was developed at Google, chose to use the Go programming language, also designed at Google \cite{Honeypot2022-yv,pike2012go}. Flux also provides a rich landscape of language bindings, one of which is Go. These shared interfaces and modularity make convergence possible.

 Given the overlapping need to schedule jobs, the first work in this space was to integrate the Flux scheduler ``Fluxion'' as a plugin scheduler for Kubernetes called ``Fluence" \cite{misale2021towards}. The rationale for this early work was that Flux could benefit Kubernetes in several ways. Firstly, Flux represents and schedules resources via directed graphs, which is notably different from the default Kubernetes scheduler that selects work for nodes based on a feasibility score \cite{kubernetes-scheduler}. In fact, Flux was created to address significant limitations of traditional HPC resource managers and schedulers by facilitating workload portability, handling high throughput job requests, and employing sophisticated techniques for flexible and fine-grained task placement and binding \cite{AHN2020202}. Enabling the Flux scheduler in Kubernetes would bring this same graph-based and hierarchical resource-aware approach to Kubernetes, and this early work demonstrated exactly that -- improved performance against the default Kubernetes scheduler~\cite{misale2021its}. More efficient scheduling was demonstrating by enabling MPI-based CORAL2 workloads to run and scale in Kubernetes that, by way of Fluence, avoided pathological resource mappings and resource starvation \cite{milroy2022one}. This work also demonstrated a valuable point -- that the scheduling provided by a workload manager must be able to concretely meet the resource demands of a workflow, but to do so efficiently and effectively to maximally utilize a set of computational resources.

Aside from the technological benefits that might come from convergence of these two specific technologies for end- and developer- users, enabling Flux to run in the cloud would also provide benefits for cloud vendors attempting to attract a larger HPC customer base. Current products that target HPC researchers \cite{aws-batch,aws-parallel-cluster,google-batch,azure-cycle-cloud} arguably serve as training wheels to help with a transition to the cloud. Other products that do not deliver a familiar command line interface would require an on-boarding process. The realization that workflows could be seamlessly portable by way of Flux, and Flux could serve as a vehicle for the workflow user-base to move between cloud and HPC, inspired the next round of work discussed in this paper. By making the full Flux workflow manager, with all components assembled, available in Kubernetes, workflow specifications that work on HPC with Flux would also work in a cloud environment also with Flux. For cloud vendors, the HPC user base could more easily make a smooth transition to using cloud too.

This paper introduces the Flux Operator \cite{the-flux-operator}, the next step in work to explore integration of a traditional HPC scheduler within Kubernetes. The Flux Operator is a Kubernetes operator \cite{dobies2020kubernetes} that handles all the setup and configuration of a fully fledged Flux cluster inside of Kubernetes itself, allowing the user to efficiently bring up and down an entire HPC cluster for a scoped set of work that optimizes for important aspects of HPC. This paper first reviews the design and architecture of the Flux Operator (Section \ref{sec:architecture}), discussing Kubernetes abstractions for efficient networking and node setup. Discussion then moves into how these design decisions impact essential needs such as workflows that use message passing interfaces (MPI), and experimental features like scaling and elasticity (Section \ref{sec:features}). Finally, experimental work shows the Flux Operator  having superior performance over the MPI Operator, the primary available option at the time for running MPI workflows (Section \ref{sec:experiment}). The paper concludes with discussion for anticipated future work, considerations for workflow design, and vision for the future (Section \ref{sec:discussion}).

% \section{Architecture}
\section{Architecture}
\label{sec:architecture}

This section details the architecture of the Flux Operator, first describing the design and needs of Flux, and how those are satisfied in Kubernetes. From the standpoint of a software architect, the task of designing the Flux Operator could be approached as a problem of pattern matching. Knowing that Kubernetes provides a set of components \cite{noauthor_undated-iq} and application programming interfaces \cite{Changes_undated-sg}, a key challenge was to assemble the components in a way that would deploy the full Flux Framework stack running inside of Kubernetes. An ideal design might aim to achieve the keystone properties of Kubernetes applications, including but not limited to fault tolerance, load balancing, and elasticity \cite{noauthor_undated-dw}. Abstractions for storage \cite{noauthor_undated-mc}, networking \cite{noauthor_undated-fw}, and workloads \cite{noauthor_undated-ud} could be selected for this design, and with a mindset of portability, meaning that the software components would be in containers themselves \cite{noauthor_undated-rz}.  
The following sections refer to two roles -- an operator developer, or someone that designs and implements the Flux Operator itself, and an operator user, an individual that installs the operator and uses it.
These architecture sections start with an overview of Kubernetes Operators, and then describe each component of Flux, and a mapping from traditional bare metal solutions to abstractions in Kubernetes.

\subsection{Kubernetes Operators}

While individual components such as pods or services \cite{noauthor_undated-ud,noauthor_undated-fw} could be individually implemented and created in the Kubernetes cluster, the advent of programmatic operators \cite{noauthor_2022-jt} in 2016 has hugely simplified this process for the developer user. A Kubernetes operator serves as a controller for one or more Kubernetes objects, meaning that a developer can express all of the custom logic needed for an application of interest in code that is compiled and run in Kubernetes \cite{noauthor_undated-bx}. The operator implements a custom resource whose behavior is dictated by a custom resource definition (CRD), a YAML file with a specification of variables for the controller to use \cite{custom-resources}. For the Flux Operator, this custom resource is called a ``MiniCluster'' \cite{noauthor_undated-tr}.  The basic design of a controller is a loop, running a reconciliation process until a cluster reaches a desired state requested by a user via this custom resource definition YAML specification \cite{dobies2020kubernetes}.  This is a declarative model \cite{declarative-programming} in that the operator user can specify high level constructs such as the cluster size and application to run, and they don't need to know the details of setting up a Flux cluster, nor do they need to consider orchestration or update of components. This approach is advantageous in that it exposes only the amount of detail that is needed for some number of jobs, and the complexity that would require niche expertise is hidden.

\subsection{Flux Framework}

Flux Framework is a novel, graph-based resource manager that is typically deployed on-premises at HPC centers \cite{AHN2020202}. It won an R\&D 100 award in 2021 \cite{Heney_undated-nq,Livermore_undated-bn}, and is currently stated to be the primary scheduler for the upcoming exascale-class system ``El Capitan'' at Lawrence Livermore National Laboratory \cite{Trader_undated-kt}.  It is called a framework because several projects combine together to form the resource manager known as Flux. A summary of these projects is described in Table \ref{tab:table}, and the interested reader is directed to the learning guide \cite{noauthor_undated-zc} for an in-depth overview.

\begin{table}[H]
 \caption{Flux Framework Projects}
  \centering
  \begin{tabular}{ll}
    \toprule
    \cmidrule(r){1-2}
    Project     & Description    \\
    \midrule
    flux-core & core services    \\
    flux-pmix     & flux shell plugin to bootstrap OpenMPI v5+   \\
    flux-sched     & Fluxion graph-based scheduler  \\
    flux-security     & security code       \\
    flux-accounting     & user bank and accounting  \\
    \bottomrule
  \end{tabular}
  \label{tab:table}
\end{table}

While Flux can be described in terms of its modules or components, for the work here it will described as it is seen in the space of Kubernetes abstractions.

\subsubsection{A Flux MiniCluster}

\paragraph{The Node}

Flux is typically deployed across nodes on an HPC cluster, where each node can be thought of as an addressable compute or storage unit, and as having a unique network address. Moving into the space of cloud, the physicality of the server goes away, and instead the basis of a node is a virtual machine. However, while Kubernetes itself is deployed on nodes, the notable object is the pod -- an abstract slice of a node that is tasked with some unit of work to run one or more containers, and allocated a particular set of resources \cite{pods}. Since the Kubernetes scheduler has no issue slicing up a single node into many pods, the first task in defining this cluster was to ensure that there was a mapping of one pod per actual physical node. The reason for this mapping is due to Flux not being able to detect running on a partial node, which is a result of its use of the portable harware locality (hwloc) \cite{hwloc} library to discover resources. The hwloc library can only detect the resources of an entire host \cite{ibm-docs-hwloc}, which in the context of two pods running on one node, would erroneously discover the same set of resources twice, double what is actually available. In practice, this would mean that Flux could schedule too much work on a single physical node that, to the resource manager, is seen as two separate nodes with identical resources. The 1:1 mapping of pods to nodes was originally achieved by way of a resource specification, a strategy that required the user to ask for just under the upper limit of CPU and memory offered by their cloud instance of choice \cite{resources}. This strategy was later improved to not require user expertise by way of rules for pod affinity and anti-affinity. These are essentially constraints that tells the scheduler to ensure  one pod per node, each with a hostname for Flux \cite{affinity}.

\paragraph{The Cluster}

While it would be possible to deploy individual pods onto nodes, early Kubernetes offered further abstractions for sets of pods such as deployments \cite{deployments} and Stateful or Replica sets \cite{statefulset,replicaset}, and in 2015, an abstraction called a Job that was the first of its kind to emulate a traditional HPC job with the intention to run and complete \cite{job-added-commit}. As of 2021, the batch working group introduced the indexed mode addition to Job~\cite{indexed-job} where the same work could be done in parallel, expecting 1 to $N$ completions \cite{jobs}. In that each node of a simple Flux cluster would be identical aside from subtle differences in startup, although other abstractions were considered, an indexed job was ultimately chosen. The indexed job is ideal in that in inherits needed features from the base Job such as having states, and adds an ability to create duplicates of the pods. To create pods it uses a batched approach \cite{batched-job-creation}, which is also advantageous to introduce an indexed ordering that ensures the index 0 is cleaned up last. This allowed us to easily design the operator to use the index 0 pod as the lead broker, and any scaling up or down of the cluster (Section \ref{sec:elasticity}) would never risk deleting the lead broker. Within this cluster context, given the assignment of one pod to one node, for the remainder of this paper the terms ``node" and ``pod" are used interchangeably as they are mapped to the same resources, memory and CPU.

% TODO HERE - this would be an easy ask from the reviewer, for us to do these actual experiments. 

\paragraph{Networking}

While it might be thought that the core of Flux is the project ``flux-core,'' one of the foundational components of Flux Framework is the scalable tree-based overlay network, or ``TBON'' that connects the core modules for scheduling and resource management. The Flux TBON is a rooted tree \cite{tbone} that features leader and follower processes (brokers), each of which is typically mapped to one node. The leader expects follower brokers to connect to it. Mapping this design to the indexed job, the role of lead broker can be assigned to index 0, and the follower brokers to indices 1 through N. Along with being easy to remember, this design decision allows pods to be created in order of their index with the lowest first \cite{indexed-job-enhancement}, and this setup is ideal to have the lead broker up earlier for the follower brokers to find. The initial networking of the cluster is done with ZeroMQ \cite{hintjens2013zeromq}, where the follower brokers identify their place in the cluster by way of their rank in a shared system configuration file, and then connect to the lead broker on a specific port via transmission control protocol (tcp) \cite{tcp}. If the lead broker is not up first, while the worker will wait to try connecting again, by default the ZeroMQ library falls back to a tcp default retry timeout that increases exponentially \cite{zeromq-docs}. In practice this means delayed cluster startup times waiting for the follower brokers to retry.  The scheduler and resource manager combined with this set of brokers that can communicate to run jobs is called a Flux instance \cite{glossary}. A Flux instance can be on the system level, meaning shared by many users, or owned by a single user. In both cases, the Flux instances handle user requests for submitting or otherwise interacting with jobs. The instance itself is hierarchical because it can spawn sub-instances whose resources are a subgraph of their parent.  

The above networking setup can give each pod a unique address that can be written into the Flux system configuration, and used to identify lead and follower brokers. For the first naive implementation, the Flux Operator created the pods, retrieved the IP addresses after creation, and then added corresponding entries to the ``/etc/hosts'' file for DNS resolution. Automated management of the hosts file proved to be a bad design because it required restarting all the pods. Instead, a later design created a headless service for the MiniCluster \cite{headless-service}, meaning that each pod could be given a label, a key value pair, that was known to the headless service, and discovering the labeled pod would add it to the network provided by the service. The headless service can then also assign a predictable hostname, which is essential for Flux to identify it. This simplified the creation of the cluster, and allowed for having the networking ready as soon as the service object was ready. Once this is done and the brokers have connected over TCP, further communication for the overlay network is done via ZeroMQ sockets \cite{AHN2020202}. However, for the cases of workflows that use a message passing interface (MPI)~\cite{mpi}, Flux has built-in modules for MPI and communication, meaning that Flux simply uses standard MPI libraries that can rely on sophisticated networking fabrics or other high speed interconnects~\cite{flux-mpi}. This reliance on cloud hardware has proven to be a challenge when deploying the Flux Operator to different cloud providers, and is a focused area of collaborative work.

\paragraph{Volume Mounts}

In terms of configuration, Flux requires a system configuration file, a ZeroMQ CURVE certificate used to encrypt communication \cite{curve}, and some means to start the brokers. In a traditional HPC setup, this means could be using a systemd service \cite{systemd}, however in a Kubernetes environment with containers, it means a start command for the container with conditional logic for the lead vs. follower brokers. In Kubernetes, all of the above configuration can be achieved via volume mounts provided via ConfigMap \cite{configmaps} objects. By way of mounting configuration files and other needed files to each pod container as read-only volumes, all nodes in the cluster have access to them.  These are mounted at \texttt{/etc/flux} and \texttt{/flux\_operator} for configurations and the starting script, respectively, and the choice of a root path affords discoverability. 

The curve certificate presented a bootstrapping design problem, as the standard way to generate it is usually via Flux itself (ZeroMQ is compiled within and exposed via the \texttt{flux keygen} command). However, this content was also required to exist for the read-only volume before starting the pod container. For the earliest design of the Flux Operator, a one-off certificate generator container was brought up that ran this key generation command, and the key was printed to the log to be retrieved by the operator. It could then be straightforward to write into the ConfigMap to be shared by the MiniCluster pods. In a later design, by way of collaboration with authors of this paper following Kubecon Amsterdam '23 \cite{CNCF_Cloud_Native_Computing_Foundation2023-pi} this bootstrapping problem was further improved by compiling ZeroMQ directly into the Flux Operator, and using cgo~\cite{cgo} to interact with it directly to generate the certificate content for the ConfigMap inside the operator. This removed an entire step to generate the one-off pod, and is a beautiful example of how sharing ideas and collaboration can lead to improvements in design and functionality.

\paragraph{A Flux Container}

The libraries and software needed for Flux along with configuration steps must happen in a common Flux container that is replicated by the indexed Job. This container would need to come pre-built not only with Flux and needed modules, but also with any application of interest to be run by the user. This is a design flaw in that most containerized applications for HPC have not been built with Flux, and would need to be built again. While the HPC community is attuned to building and optimizing components for different architectures, ideally a more modular, cloud-native solution would not require investing time to do that. Once required software and configuration files are present, setup continues  to create either a single-user or site-wide installation of Flux. The Flux Operator opts for a single-user design, and enables customization via variables exposed on the CRD. This customization includes (but is not limited to) archiving of data, creating multiple users, starting in an interactive mode, starting a RESTful application programming interface, or creating a custom batch job. The final component of the container is the ``entrypoint" or the command that is run when the container is started. This varies between the lead and follower brokers, where the lead broker typically starts with a command to run a job, and the follower broker starts expecting to connect and receive work.

% \section{Features}
\section{Features}
\label{sec:features}

Once it was possible to run and complete a basic workflow (discussed in Section \ref{sec:experiment}), development thinking moved toward adding desired features for such a workload manager in Kubernetes. This section will describe early work to enable scalability and saving state, elasticity and auto-scaling along with workflow integration. These features for workflows, along with the core design of the Flux Operator, are considered experimental in the sense that they are implemented with the goal of testing and improvement in mind. The below represents a sample of this work, and more experiments can be found in the examples directory of \href{https://github.com/flux-framework/flux-operator}{https://github.com/flux-framework/flux-operator}.

\subsection{Saving State}

The goal of experiments to save state would involve starting a Flux MiniCluster, running some number of jobs, pausing them, saving the state of the job queue, and then bringing it down to bring up a differently sized cluster (larger or smaller) to load the jobs into, where they would continue running. This concept of saving state is similar to forensic container checkpointing \cite{forensic-container-checkpointing}, an experimental idea for Kubernetes, and would be useful for pausing workflows for cost savings or waiting on resource availability. These experiments varied based on when the queue was paused. In the earliest tests, job completion was required before saving state, while for later tests, jobs were stopped mid-run. 

In practice saving state meant waiting for the queue, pausing, and then saving to an archive in a shared volume between two MiniClusters. The first MiniCluster pods were then waited for to terminate, and the new MiniCluster was waited for to come up and restore the jobs. It was observed that jobs would successfully save and load into the new cluster, maintaining job identifiers and size, however when stopping a running queue, 1-2 jobs could be lost between the transfer. While the reason for this loss would be interesting to understand, as it is an experimental prototype for a feature, the work is beyond the scope of this paper, and akin to other features discussed here, should be pursued with a compelling research use case. When this time comes, more analysis would be needed to understand exactly what is going on. The majority of jobs (e.g., a rough estimate of 9 out of 10) transition nicely, meaning that a job on the previous queue can get scheduled to a new larger or smaller cluster. As would be expected, if a job is moved onto a cluster lacking enough resources, it would logically not be scheduleable.  A write-up and tutorial to reproduce this work is available \cite{saving-state}.

While this early work to save state was simple, it was a glimpse into the idea that scheduled workflows could in fact be moved.  In changing the size of the resources available by way of creating a new MiniCluster, it was the earliest prototype for what might be called scaling or elasticity, discussed next.

\label{sec:elasticity}
\subsection{Elasticity}

Elasticity can be thought of as automated dynamic scaling \cite{10.1145/3341325.334199}. Instead of making a cluster larger or smaller by way of saving state and loading into a different size, true elasticity means changing the size of a single cluster, which in the context of the Flux Operator means that Flux must adapt dynamically. To accomplish this, ideally Flux needed to have support for resource dynamism, however it did not. Short of that there was a way -- one that might be considered a hack -- to get this to work. The following steps enable an elastic Flux MiniCluster:

\begin{itemize}
    \item A max size variable is added, meaning more nodes are defined in the system configuration file than actually exist.
    \item Flux is told to create a cluster at a size that is between 1 and this max size.
    \item Any change request to the CRD (from a user or API) validates the request, and updates the indexed job.
    \item An update to increase in size creates new pods.
    \item An update to decrease in size terminates pods. 
\end{itemize}

The above also carries the constraints that the cluster cannot be smaller than one node (only a lead broker) or larger than the initial ``maxSize." The larger indices are terminated first and the operator does now allow reduction to size zero, so the lead broker is never at risk of deletion -- such a request would delete the entire MiniCluster that relies on it. The reason this works is that Flux sees the initial set of pods that do not exist as simply being down, which is happens frequently in a high performance computing environment. When the nodes are created their corresponding follower brokers start, ping the lead broker on the port to connect (typically port 8050) and then they seamlessly join the cluster. From the standpoint of the user, they change the ``size" value in their MiniCluster CRD, apply it, and then see their cluster grow or shrink. The Flux instance run by the lead broker simply sees a node come online. On the Kubernetes side, this ability for the indexed job to have elasticity requires a minimum Kubernetes version of 1.27 \cite{elastic-indexed-job}. 

At this point, it needed to be decided what might trigger this change in size. Elasticity was implemented in two ways, first with an application-driven approach \cite{basic-elasticity-three-bears} that required extra permissions to be given to the in-cluster service account, allowing the application inside the cluster to ask for more or fewer pods directly. It was then discovered that Kubernetes has autoscaling APIs intended for this use case. This autoscaling approach is discussed in the next section. 

\subsection{Autoscaling}

In Kubernetes there are two types of scaling - horizontal and vertical. Horizontal typically refers to adding pods, while vertical refers to adding resources to existing pods \cite{hpa}. Both are based on the idea that resources should change in response to changing workload needs -- if cluster or resources are too big, they are made smaller, and vice versa. In the case of the Flux Operator the primary interest was horizontal auto-scaling, or changing the number of pods to dynamically increase or decrease the size of the MiniCluster to respond to the demands of a workload. This led to a first try to implement horizontal pod auto-scaling (HPA)~\cite{hpa} using the HorizontalPodAutoscaler API resource, a cluster API to watch a selected set of pods for resource consumption and increase or decrease the number depending on utilization. For the simplest cases, a default autoscaler was first deployed that considers a metric such as percent CPU usage and uses an algorithm \cite{autoscaling-algorithm} to calculate a target scale value for the number of pods. This could be tested by running a CPU intensive simulation \cite{elasticity} to observe the autoscaler adding and removing pods. However, the approach was not fine-tuned enough to the potential needs of an application being run by Flux. Instead of an arbitrary decision to add or remove pods based on CPU, a design more specific to Flux is warranted. As an example, one design might be that the Flux lead broker makes decisions about when and how to scale depending on the content of the queue. Another valid design would be to allow for changing the size of a single running job. Both of these ideas, and more generally designs for autoscaling, are valid and prime for future work.

With this in mind, a custom metrics API was implemented \cite{VanessaSaurus2023-ce}, meaning implementing an equivalent API endpoint controller that would be called by an autoscaler with instructions for how to scale the cluster. This resulted in the Flux metrics API \cite{flux-metrics-api}, a standalone API that runs directly from the lead broker pod and provides decisions about scaling up or down based on the size or other metrics about the queue. With this API, it was possible to demonstrate an auto-scaling operation running based on a trigger coming directly from Flux. More work will be needed to test this setup with real workflows. In the meantime, more details about this setup and basic elasticity are available in an external post \cite{elasticity}.

One notable feature about the implementation of the autoscaling approaches described above is that regardless of whether the request comes directly from a user changing a value in a file or an application or a programmatic autoscaler, the same internal logic (functions) are used to validate and then perform the patch.

\subsection{Multi-Tenancy}

Multi-tenancy refers to the ability to support multiple users on the same resources. This is not a common design in Kubernetes, as ownership of resources is typically designated by namespaces, custom permissions on connected resources like storage, and role based access controls (RBAC) \cite{multi-tenancy}. Recognizing these challenges, as an early approach there are several modes of interaction:

\begin{itemize}
 \item Single user: the user owns an entire MiniCluster, and uses the default Flux user in the container
 \item Multiple users: controlled via PAM \cite{PAM} authentication
 \item Multiple users: controlled via RESTFul API access
\end{itemize}

In anticipation of the last two cases that implement multi-tenancy, a RESTFul application programming interface (API) \cite{flux-restful-api,flux-restful-api-zenodo} was designed that runs from the lead Flux broker pod, and thus exposes interactions with Flux to submit, get info for, cancel, and otherwise interact with jobs via Flux Python bindings exposed by the API. This is made possible by exposing the internal port that the API is running on via an external NodePort \cite{node-port-service} and port forwarding \cite{port-forward} an external client outside of the cluster can interact with it. 

In all cases of requiring authentication, the Flux RESTful API uses an OAuth2 based approach, storing a database of user identifiers and encoded passwords \cite{fop-users}, and first authenticates by using a base64 encoded username and password (typical of a basic authentication scheme \cite{basic-http}), and then provides the user with an expiring token that can be used for further interaction. In the case of a single Flux user behind a multi-tenant API, this authentication and authorization happens, and then all users submit jobs to a shared queue. In the case of more true multi-tenancy with PAM, the custom resource definition asks for usernames (and optionally, passwords) in advance, and then creates the accounts on the system that are checked after authentication. The installation of flux-accounting \cite{flux-accounting} can then be enabled for the lead broker's queue, and use a traditional fair-share algorithm to determine job priority \cite{flux-accounting}. This work can be extended with more cloud-native approaches that take advantage of namespaces and roles, such as is described later (Section \ref{sec:workflows}).

\subsection{Bursting}

To complete the early work in autoscaling, the concept of bursting was considered \cite{drako2022need}, which means not just extending the size of a local cluster, but actually extending work to external resources. The bursting work for Flux would extend this approach to not just deploy external resources, but allow the lead broker to connect to brokers that are deployed in the other clusters. As an example, a Kubernetes cluster running on Google Cloud might burst to a cluster running on Compute Engine (CE), or to a cluster on Amazon Elastic Kubernetes Service (EKS). 

To implement a prototype for bursting, a simple design was chosen first. A plugin service would be running from the lead broker of a primary cluster, and the running user would load one or more bursting plugins into it. Each bursting plugin is targeted to a particular provider (e.g., EC2 or CE). While there are many ways to trigger a burst, a simple approach of looking for an attribute ``burstable'' on a job set to true was chosen first. This request could be done on the command line. Upon discovery of this attribute, the bursting service attempts to schedule the job with the plugin. Each plugin is free to decide if the request is satisfiable by its own custom terms. If the burst is satisfiable, the job is assigned to the bursting plugin, and the plugin creates a new cluster or assign the job to an existing cluster. In the case of creation, the technique of telling the primary cluster that there are more nodes that are expected (and start down) than there actually are is used, and assign them namespaced hostnames that will correspond to the bursted cluster. The calls that are necessary to bring up the second cluster are run, which might mean deploying Terraform \cite{the-non-tofu} configuration files or creating a second Kubernetes cluster via API calls, and then the cluster starts just as a local MiniCluster would. The key difference, however, is that the lead broker of the primary cluster is exposed as a NodePort \cite{node-port-service} service that can be discovered by the external cluster. The secondary brokers, all followers, then come up, find their hostnames in the ranked system configuration, and connect to the lead broker IP address from another cluster. To the user, they simply see that the nodes are down, and then they come up when the cluster is ready. Jobs that are scheduled on the primary broker queue that possibly could not run due to insufficient resources can then run. At the time of this writing, the main bursting service is implemented along with four bursting plugins for each of GKE, EKS, CE, and a local burst \cite{flux-burst}. 

Finally, the bursting service is designed to be modular and flexible. Aside from being able to load different plugins, it allows for customization of the function provided to select a burstable plugin, to interact with the queue, and to select jobs. A mock version of a Flux job is also available for development. The work in bursting is still early, and akin to elasticity, work on these prototypes should continue to eventually develop more hardened Flux modules and algorithms for bursting.

\subsection{Workflow Integration}
\label{sec:workflows}

Running a single MiniCluster to create an isolated Flux cluster in Kubernetes is a good first step, but not sufficient for real-world use cases of complex workflows. While it would be possible to shell into or otherwise interact with the cluster and run a workflow tool that implements Flux as an executor \cite{snakemake,nextflow,streamflow}, this also does not enable features needed for complex, heterogeneous workflows that might require different sizes or configurations of MiniClusters. For this reason, the authors of this paper have started to think about how to integrate the MiniCluster custom resource definition as a first-class citizen into workflow tools.

After the Kubecon Amsterdam '23 presentation, collaborators (including author AC) were quickly motivated to add the Flux Operator as a job type to the workflow tool Kueue, a Kubernetes-native job submission operator that handles managing multi-tenancy of Kubernetes batch jobs that can be produced by a number of operators \cite{kueue}. A similar approach is being developed to have a workflow tool control creation and management of MiniCluster, an idea being implemented into the Snakemake \cite{snakemake} workflow tool as a Kueue executor plugin \cite{snakemake-executor-kueue}. Defining even an example workflow for high performance computing is a non-trivial problem, as many codes are either private or not portable. This initial work with the Flux Operator is hopefully setting the stage for doing more exploratory work for integrations of this type. Tackling this early problem is a two-fold challenge to design technologies and inspire more collaborative opportunities for the HPC community. 

% \section{Experiment}
\section{Experiment}
\label{sec:experiment}

The Flux Operator is compared to another state-of-the-art Kubernetes Operator, the MPI Operator, which at the time of the experiments was considered the main option in the field for running MPI workloads \cite{mpi-operator-openshift,biliaiev2021enabling}. The MPI Operator started as part of the Kubeflow project and defines an ``MPIJob'' custom resource \cite{mpijob}. Unlike the Flux Operator that coordinates between brokers with ZeroMQ, the MPI operator coordinates workers via secure shell (SSH) \cite{mpijob-ssh}. It requires an extra launcher node that serves the sole purpose of coordinating the workers, and akin to the Flux Operator, uses dedicated hostnames with an equivalent headless service. This launcher node conceptually could be thought of as analogous to the lead broker of the Flux instance in that it serves as an entrypoint for submitting jobs, however the main difference is that the Flux lead broker is considered part of the cluster to perform work. The MPI launcher node is not, and in practice this means the user will need to always incur the costs of an extra node just for the launcher.

\subsection{Methods}

\begin{figure}[H]
  \centering
  \includegraphics[scale=0.3]{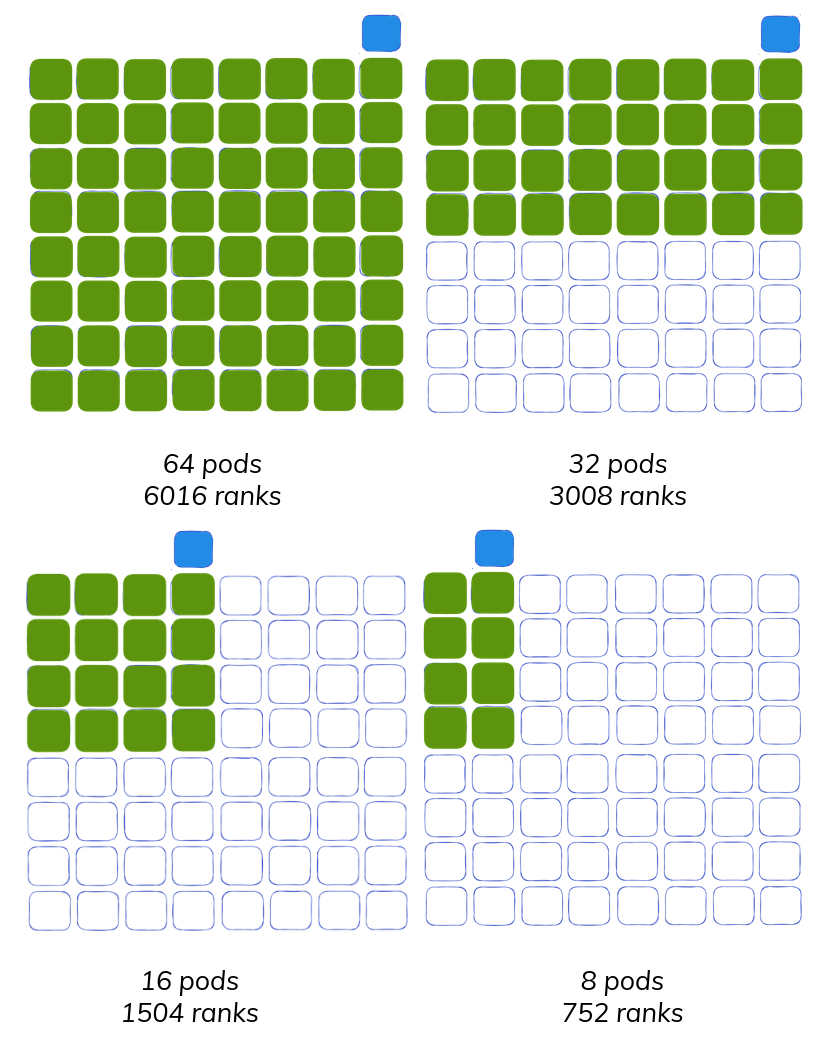}
  \caption{Experiment design for testing the Flux Operator against the MPI Operator. A single Kubernetes cluster of size 65 was created (blue outline) to test subsequently smaller cluster sizes, including 64 pods (3008 ranks), 32 pods (3008 ranks), 16 pods (1504 ranks), and 8 pods (752 ranks). An extra node (blue) is needed for the MPI Operator launcher to supplement the nodes doing work (green).}
  \label{fig:fig1}
\end{figure}

Experiments were conducted on Amazon Web Services Elastic Kubernetes Engine (EKS) \cite{eks}, using the \texttt{hpc6a.48xlarge} instance type \cite{hpc6a} with the elastic fiber adapter (EFA)~\cite{efa}. intending to test the Large-scale Atomic/Molecular Massively Parallel Simulator (LAMMPS)~\cite{LAMMPS} in a strong-scaling configuration across cluster sizes and ranks of 64/6016, 32/3008, 16/1504, and 8/752, respectively (Figure \ref{fig:fig1}). This design was chosen to mirror previous work \cite{misale2021its}. LAMMPS is used by the Collaboration of Oak Ridge, Argonne, and Livermore (CORAL) as a representative scalable science benchmark as part of CORAL-2~\cite{coral2-bench}. In testing, LAMMPS runs in parallel on MPI ranks (processes) across nodes, a molecular simulation \cite{Van_Duin2001-nw} and problem size $64 x 16 x 16$ was chosen that would adequately test strong scalability across the chosen rank and node counts. LAMMPS scalability depends on network latency, and the experiment results report the total wall time recorded by the LAMMPS run as a metric for performance. A cluster setup that enables lower latency will run faster, and ideally the simulation should get faster as the number of nodes is increased with strong scaling. A second metric time of interest is the time for the launcher to submit and complete a job. For Flux this means timing the \texttt{flux submit} command that is given the command to run LAMMPS, and for the MPI operator it means timing the \texttt{mpirun} command that does the same. The final metric time of interest was the total cluster creation and deletion times, which can be calculated based on the total runtime of the MiniCluster minus the LAMMPS total wall time. This time would include each pod preparing the broker, starting Flux, and networking with the lead broker. The runtime would ideally decrease across these chosen rank and node sizes. 

To ensure the nodes of the cluster are consistent and do not influence results, experiments were run on the same Kubernetes cluster, and simply used smaller portions of it. As the MPI Operator requires an extra launcher node, the maximum cluster size needed (64) was increased by 1, resulting in a size 65 node cluster for these experiments. Finally, a modified version of the MPI Operator was used \cite{milroy2022one} to allow it to scale to over 100 MPI ranks.

The experiments proceeded as follows. The main Kubernetes cluster of size 65 is first created. Then, for each of the Flux Operator and MPI Operator:
\begin{enumerate}
    \item Launch Job / create MiniCluster for sizes 64, 32, 16, 8
    \item Run LAMMPS x 20
    \item Record timings and save all configuration files and logs
\end{enumerate}

For each experiment run, a single ``throwaway'' run is first performed to pull the container with Flux and LAMMPS to the node, where it is cached for further runs \cite{images}. This ensures that time recorded in creating the MiniCluster does not include pulling the container image, which would have variability depending on the image size.  The experiments are then run in an automated fashion using Flux Cloud \cite{flux-cloud}, a simple experiment orchestration tool for running experiments with Flux MiniClusters on Kubernetes.  All experiment code, configuration files, and tagged containers \cite{flux-operator-container,mpi-operator-container} are available \cite{kubecon-run6}.

\subsection{Results}

\paragraph{Cluster Creation and Deletion}

\begin{figure}[H]
  \centering
  \includegraphics[scale=0.32]{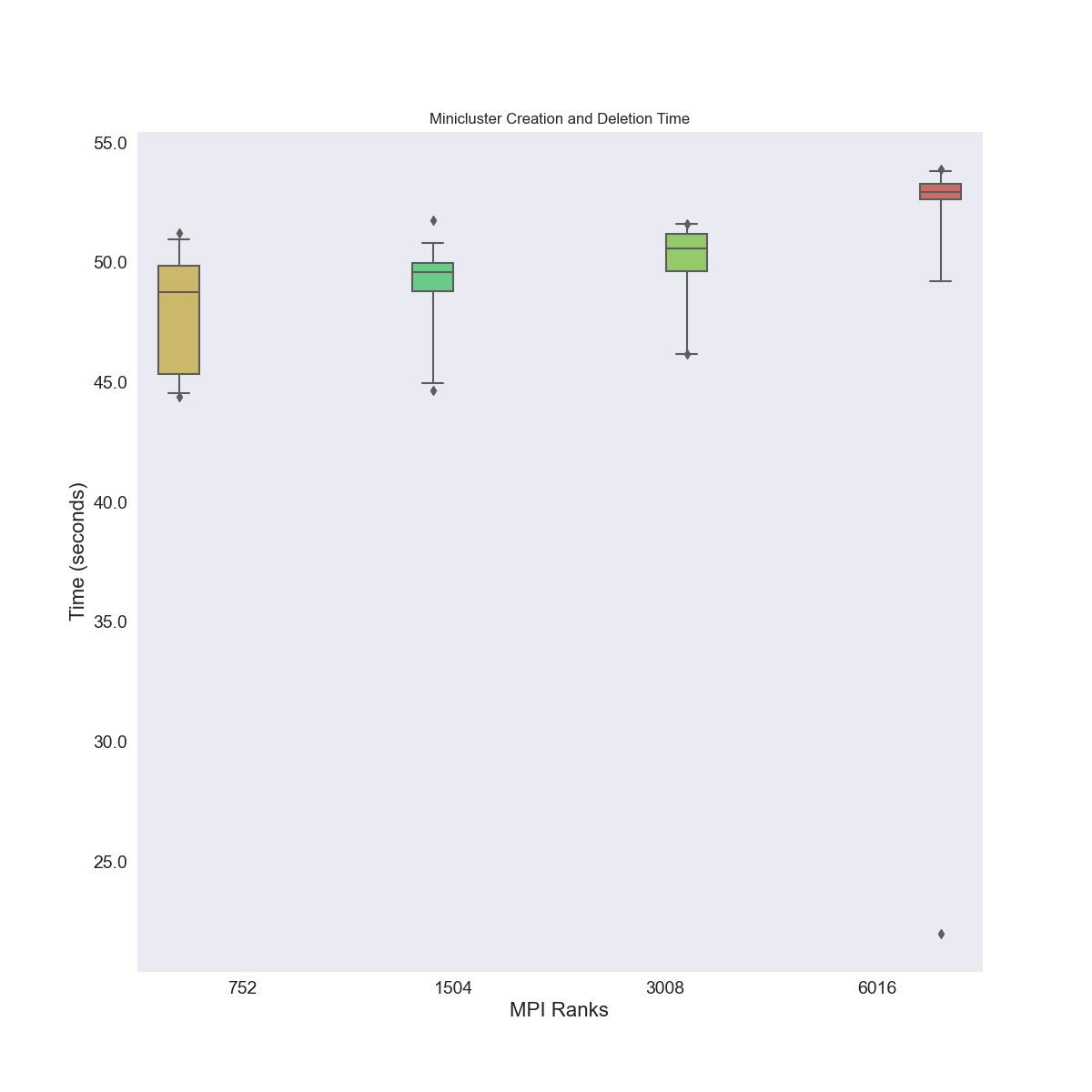}
  \caption{Combined Flux MiniCluster Creation and Deletion Times across MPI ranks, corresponding to cluster sizes of 8, 16, 32, and 64 nodes, respectively. These experiments were run on Amazon Web Services with the hpc6a.48xlarge instance type and the elastic fiber adapter for networking.}
  \label{fig:fig2}
\end{figure}

Bringing up and down clusters of sizes 8, 16, 32, and 64 across 20 runs each, a weak linear scaling was observed (Figure \ref{fig:fig2}) indicating that the indexed job could efficiently create pods. All sizes were created and ready in under a minute with variability of approximately 5 seconds. Likely this variability reflects the slowest or last node to come up, as the cluster is not considered to be completely up until all nodes are ready. This result was surprisingly good, as it could have scaled differently.
% Not sure how to describe "differently"

Due to the design of the MPI Operator, there was not a corresponding measure of creation time appropriate for comparison, so the two are not comparable side by side.

\paragraph{LAMMPS Total Wall Time}

\begin{figure}[H]
  \centering
  \includegraphics[scale=0.24]{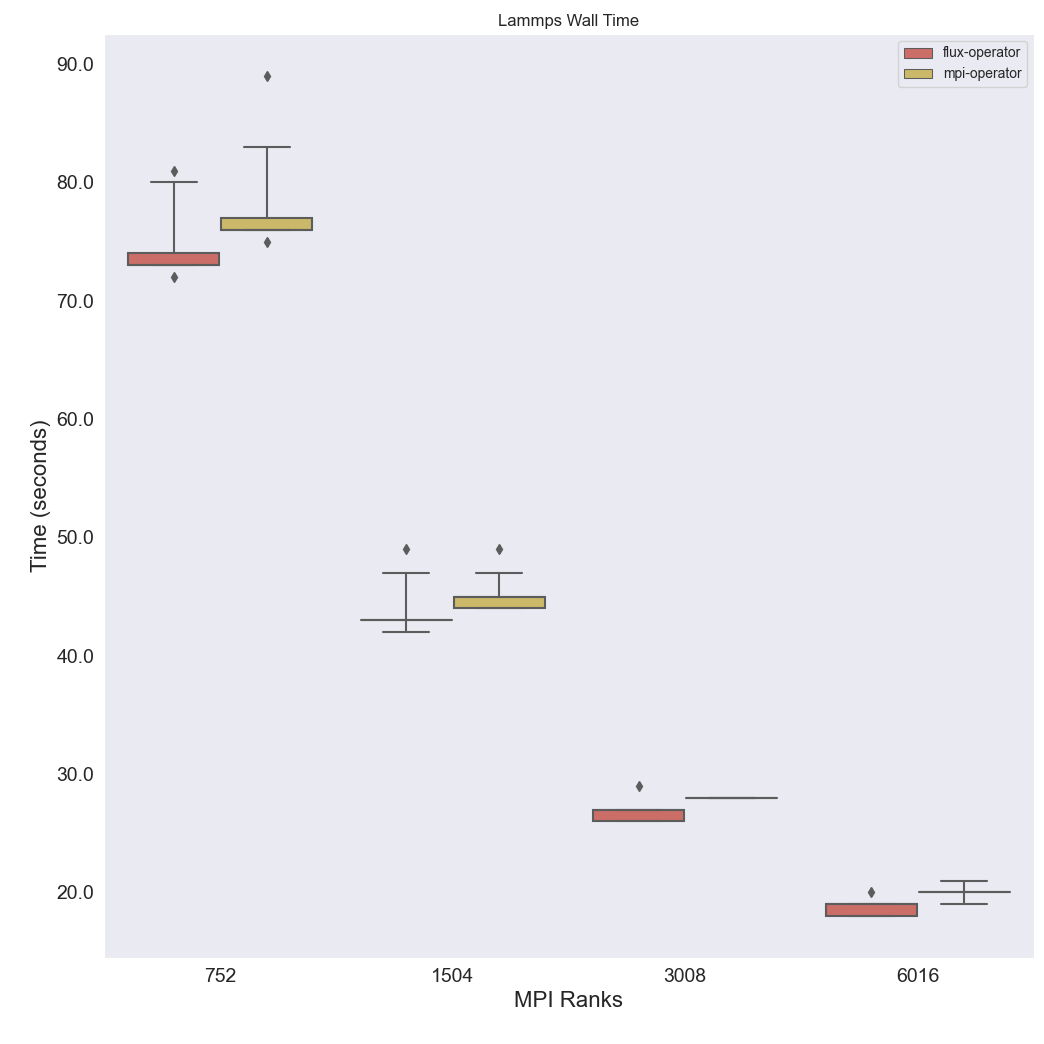}
  \caption{Total wall time, as reported by the LAMMPS software, between the Flux Operator and MPI Operator. The Flux Operator is consistently faster, a result that could be more impactful for longer running experiments.}
  \label{fig:fig3}
\end{figure}

A primary time of interest for running LAMMPS is the total wall time, which is reported by the LAMMPS software itself. Consistently better performance of the Flux Operator over the MPI Operator was observed, with mean times that are approximately 5\% faster, meaning that the Flux Operator software completed the same workload more efficiently (Figure~\ref{fig:fig3}). While the differences in means for this experiment are small, this improvement would likely be more prominent for longer experiments, and could lead to reduction in total costs. Identifying the underlying reasons for the improved performance is another task suitable for investigation with performance tools in future work.

\paragraph{Launcher Times}

Comparing launchers (\texttt{flux submit} for the Flux Operator, and \texttt{mpirun} for the MPI Operator) there is a slightly larger time difference (Supplementary Figure \ref{fig:fig5}), where both generally perform well under strong scaling (the time goes down). What is unknown is whether there is an inflection point at larger scales where the MPI Operator might plateau or otherwise show a different pattern. The resources were not available to run these larger experiments at the time, but these patterns and scaling can be explored in future work.

\paragraph{Design Considerations}

\begin{figure}[H]
  \centering
  \includegraphics[scale=0.2]{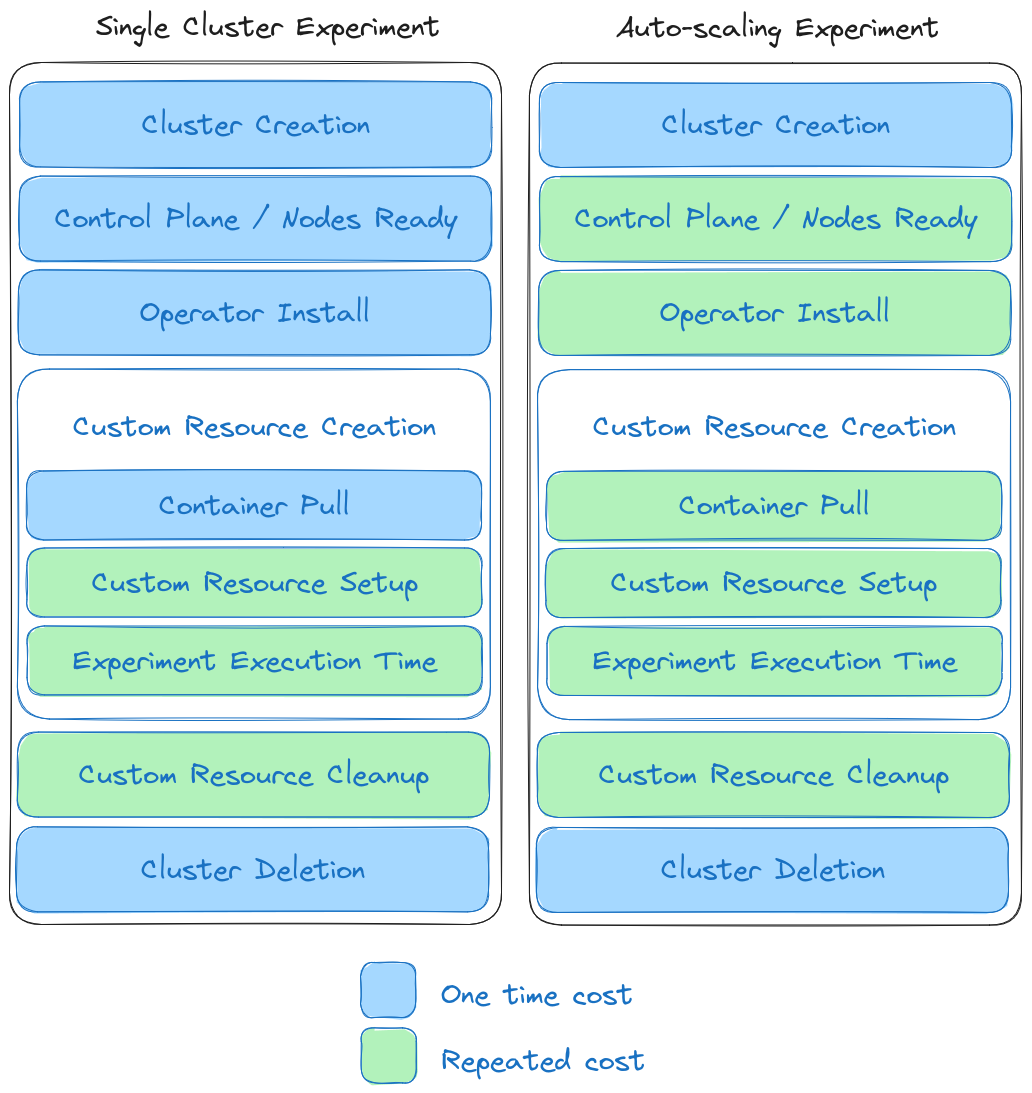}
  \caption{Times to consider when performing experiments with a Kubernetes operator, from the point of creating the cluster to deleting it. For a single cluster without autoscaling (left) many of the operations are one time costs (blue), while for an auto-scaling cluster that needs to provision new nodes and pull containers to them, the costs become repeated (green).}
  \label{fig:fig4}
\end{figure}

\begin{figure}[H]
  \centering
  \includegraphics[scale=0.25]{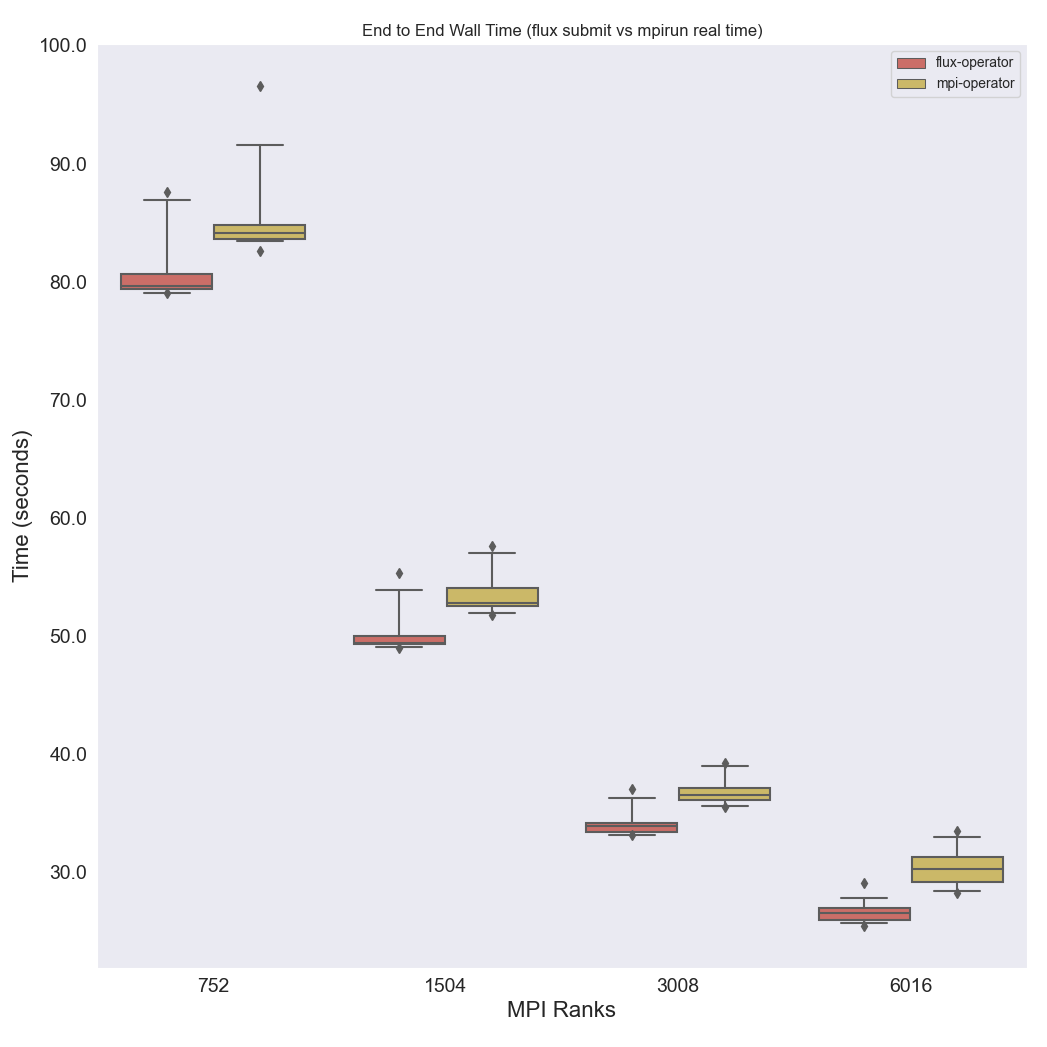}
  \caption{Launch time measuring flux submit (the Flux Operator) or mpirun (the MPI Operator). The Flux Operator is consistently faster, a result that could be more impactful for longer running experiments.}
  \label{fig:fig5}
\end{figure}

Anticipating interest in running experiments of this type, where there is generally some operator in Kubernetes that is going to pull one or more containers to Kubernetes and perform scoped work, a visualization of the steps that have salient times are provided, and the the interested reader is encouraged to think about them (Figure \ref{fig:fig4}). 

% https://excalidraw.com/#json=aDCm6xjnA1n9xeQ5N82SU,yTBSPcWtxxEnQFqHdk5HNQ

In Figure \ref{fig:fig4} above, there is a distinction between an operator setup that is using autoscaling (right) vs. not (left), and costs that are incurred once (blue) vs. repeated (green). Cluster creation generally means the start and setup of instances, along with any networking and devices that are needed. The primary difference between the two scenarios is that an autoscaling cluster is going to be adding new nodes, meaning that the cluster will need to provision those nodes, and the user will be required to wait. Thus, this update process for the cluster that requires the user to wait (and pay for the time) becomes a repeated cost. This also means that a typically one time cost of pulling a container may occur several times, primarily when new nodes are added. Note that this diagram assumes experiments running on one cluster. An autoscaling setup that employs bursting to new clusters would need to consider the additional time of creating and deleting the new clusters.

It is suggested to the reader to consider these times, along with the differences between a setup with autoscaling versus one without or potentially bursting, for future experiments when anticipating costs. Notably, the setup time for any particular operator could be generally consistent, and variance has been seen (but is not reported here) in the other steps between cloud providers. A more critical study and understanding of these times is warranted for future work.

% \section{Discussion}
\section{Discussion}
\label{sec:discussion}

This work demonstrates improved performance using the Flux Operator for running an HPC workflow in a cloud-native environment. This early innovation comes with strengths, limitations, and and desires for future work that include the topics of workloads and scheduling, storage, tenancy, and cost-estimation, among others. 

A discussion of limitations and further hopes for innovation is needed for transparency of this work. First, it is a design flaw that the main execution container is required to have Flux and the application of interest. This means that any user of the Flux Operator is required to rebuild their container in full to include Flux, which not only requires the work to do the re-build, but some limited knowledge of Flux. While container requirements are provided alongside the documentation \cite{container-requirements}, this is not good enough. While the dependencies and complexity exist to enable advanced capabilities, the authors believe there are approaches that can improve upon this strict requirement.

A next limitation is the creation of the entire MiniCluster using a single indexed Job. While this is the ideal for the time being, as the indexed Job is released with core Kubernetes, an eventual refactor to use a JobSet \cite{jobset} would be desired, which allows for the same indexed Job to be used, however with better ownership of assets, and an ability to define different groups of nodes each as a Replicated Job under the same network namespace. Allowing for different sets of nodes would not only make it possible to separate logic between the lead and follower brokers, but also allow for creation of a MiniCluster with different pod specifications mapped to different resource needs. JobSet would also allow for better definition of a success policy, or explicitly saying that the job is completed when the lead broker exits. Author VS has created a prototype using JobSet, and hopes to continue in this direction when it is considered for Kubernetes core.

For next steps of work for experimental features, the Flux software itself needs innovation for the set of the hacks that were implemented. The ability to scale up and down dynamically without ``registering'' the non-existent nodes in the Flux system configuration is a good example, along with a more hardened approach for bursting that likely comes down to plugins written directly in C or C++ alongside Flux. Especially the work in bursting is early and exciting, and will be continued in future work. Notably, the experiment application did not require use of storage, and while there are several tutorials and examples for different cloud solutions, this is an entire area of the design that requires further work and thinking.

Another challenge is that of poor workflow consistency and reproducibility. While there are scattered workflow tools that are used by HPC centers (e.g., national labs), these authors consider much of the HPC community behind with respect to the reproducible workflow movement. Part of this work moving forward is to not only identify proxy applications and workflows, but also to containerize them, and make it quick for an interested party to run them easily in a cloud environment. Part of this work will not only be understanding how they work in containers and across a Kubernetes cluster, but also developing means to assess performance. 

An understated challenge in the converged computing space is also culture and communication. As stated in the introduction, convincing one side to be open to ideas from the other is a non-trivial task. For basic communication, if there is discussion between HPC and cloud community members, a simple term like ``node'' or ``scheduler'' can mean something different. This might be tackled through discussion, and creation of a shared lexicon that allows for talking about comparable distractions. This introduces a further challenge when looking at the means for communication. Academic groups tend to write papers (and industry groups less so), and developing software in research is made more complex by the publication incentive structure that wants to highlight new research results \cite{Merow2023-tm}. Practically speaking, both the HPC and cloud communities will need to meet one another half way. This might mean researchers presenting work at (traditionally) more cloud-oriented conferences and venues, or cloud developers participating in more traditionally research-oriented venues. For both, it means distributing knowledge through blogs and other common mediums. This work calls out to cloud vendors an immense desire to work together. While it is  understandable that there is a primary concern about direct comparison, there is a path for respectful collaboration, developing technologies that can work across clouds, and learning from one another.

These experiments were run on one cloud with a particular networking fabric and instance type, and at a maximum scale of 64 nodes, which is very small for traditional HPC. However, the recent work to train large language models \cite{Shen2023-wi} provides a common use case for needing scaled resources, and might allow for shifting incentives toward that. One of the most challenging decision points for running experiments of this nature is the sheer size of the space of potential experiments that could be run. As the goal of this work was to compare the two operators with an HPC application, the choices made reflect that goal, and a desire to optimize performance (choosing a configuration to support low network latency) as much as possible. These same experiments run on other instance types, interconnects, or even regions could have different results. Further experiments should be pursued that continue to test scale, elasticity, and performance in the space of networking, IO, and application metrics.

The Flux Operator brings several features that could be helpful to more general Kubernetes workflows. The first is that using a Flux cluster inside of Kubernetes gets around some of the infamous etcd limits or bottlenecks \cite{Larsson2020-bu}. Submitting to Flux does not stress Kubernetes application programming interfaces or etcd, and could scale to hundreds of thousands to potentially millions of jobs \cite{Livermore_undated-bn,AHN2020202}. The second is the hierarchical way of looking at heterogeneous tasks. Kubernetes would benefit from having more flexibility about telling tasks where they can go, and then binding them to exactly the resources needed. This brings up tension between a more manual vs. automated decision made by the Kubernetes kubelet. The Flux Operator does something that is not native to Kubernetes to help this issue. By way of allocating a pod to a node and giving control to Flux, possibly ineffective bindings decisions that are made by the kubelet can be avoided. The Flux Operator allows Flux, a workload manager accustomed to making intelligent resource bindings, to take control. To step back, ideally there should (or could) be a mechanism in Kubernetes to enable more performance oriented decisions that the kubelet makes. Having a consistent view of resources that the Kubelet is exporting as the final truth via cgroups is not necessarily desirable, and there are several reasons why. The first is that there are many ways to slice up a node, and a ``best" way depends entirely on the application in question. For example, some applications may perform well given an equal split, while others  might be optimally be broken across sockets. This suggests that granularity on the level of the socket is needed, which is not currently exposed in Kubernetes. A concrete example comes from the MuMMi workflow \cite{Di_Natale2019-ks} that requires one dedicated CPU and socket to be close to a PCI express us for a particular GPU to minimize trading data back and forth between CPU and GPU memory. This level of granularity is not currently exposed in Kubernetes, nor are there efforts to understand applications on this level. This is a key area for innovation and collaboration, and understanding basic design patterns for networking, IO, and application performance is likely a good start. Ideally applications that are running in Kubernetes today could be better understood via performance analysis, and a decision made about if the time required to optimize is worth to be invested for the potential benefit.

% \section{Conclusion}
\section{Conclusion}
\label{sec:conclusion}

The popularity and economic clout behind cloud computing presents a challenge for the high performance computing community -- to resist new paradigms of cloud-native technologies and fall behind, losing talent and customers, or to embrace it, pursuing technological innovation, and viewing the shift as an opportunity for collaboration and growth. The latter approach, a movement identified as ``converged computing'' is a good path to take, and this work represents an early start towards that desired future. The work here starts with one of the highest levels for running workflows -- the orchestration tool -- and has thought about the convergence of the HPC workload manager Flux Framework with the cloud orchestration framework Kubernetes. The Flux Operator is an example of convergence of technologies in this workload management space, and has demonstrated superior performance to the current equivalent in the space. In sharing this process of thinking about design to implementation, the authors of this paper hope to start discussion with the larger community that spans cloud and HPC for how to think about working on convergence for other types of software and technology. This work is a shining example of the collaboration and fun possible. The sharing of these results at Kubecon Amsterdam '23 \cite{CNCF_Cloud_Native_Computing_Foundation2023-pi}, inspired collaboration, discussion, and excitement about the converged computing movement. Projects and work are underway to address gaps that have been identified, and each a collaboration between computer scientists and cloud developers. The authors of this paper hope that this early work inspires, and allows for continued discussed for innovation in this space for not just workloads and scheduling, but also the challenges around it -- storage, tenancy, and cost-estimation, among others. Through collaboration and creativity of design, pathways can be discovered for moving seamlessly between the spaces of cloud and HPC. Converged computing is a paradigm shift, and it's up to the HPC and cloud communtities to decide to embrace change and grow from it, or fight it off. This work chooses the first approach, and embraces it with hopes for a better future, and stronger more reproducible science.

%%%%%%%%%%%% Supplementary Methods %%%%%%%%%%%%
\footnotesize
\section*{Supplementary Figures}

%%%%%%%%%%%%% Acknowledgements %%%%%%%%%%%%%
\footnotesize
\section*{Acknowledgements}

We would like to thank the entire Flux Framework team for regular discussion on design points, their immense expertise and experience in this space, and a safe, fun environment to learn and grow. We give our gracious thank you to both Amazon Web Services and Google Cloud for their support. This project was supported by the LLNL-LDRD Program under Project No. 22-ERD-041.

This work was performed under the auspices of the U.S. Department of Energy by Lawrence Livermore National Laboratory under Contract DE-AC52-07NA27344. LLNL-JRNL-855145-DRAFT.

%%%%%%%%%%%%%%   Bibliography   %%%%%%%%%%%%%%
\normalsize
\onecolumn
\bibliography{reference}

%%%%%%%%%%%%  Supplementary Figures  %%%%%%%%%%%%
%\clearpage

%%%%%%%%%%%%%%%%   End   %%%%%%%%%%%%%%%%
%\end{multicols}  % Method B for two-column formatting (doesn't play well with line numbers), comment out if using method A
\end{document}